# Mechanical Ductile Detwinning in $CH_3NH_3PbI_3$ Perovskite

Li Yang[1], Jinjie Liu[1], Yanwen Lin[1], Ke Xu[1], Xuezheng Cao[1], Zhisen Zhang[1,*] and Jianyang Wu[1,2,*]

[1]Department of Physics, Research Institute for Biomimetics and Soft Matter, Jiujiang Research Institute and Fujian Provincial Key Laboratory for Soft Functional Materials Research, Xiamen University, Xiamen 361005, PR China

[2]NTNU Nanomechanical Lab, Department of Structural Engineering, Norwegian University of Science and Technology (NTNU), Trondheim 7491, Norway

**Abstract:** Twin boundaries (TBs) were identified to show conflicting positive/negative effects on the physical properties of $CH_3NH_3PbI_3$ perovskite, but their roles on the mechanical properties are pending. Herein, tensile characteristics of a variety of TB-dominated bicrystalline $CH_3NH_3PbI_3$ perovskites are explored using molecular simulations. TB-contained $CH_3NH_3PbI_3$ are classified into four types from their tensile ductile detwinning characteristics. Type I is characterized by smooth loading flow stress-strain responses, originating from relatively uniform stress distribution induced gradual amorphization at TB region. Types II and III are represented by sudden drop of loading stresses but then distinct ductile flow stress-strain curves, resulting from limited and large-area amorphizations of TB-involved structures, respectively. However, Type IV is highlighted by double apparent peaks in the loading curve followed by ductile flow response, coming from stress-concentration of localization-to-globalization at TB structure, as well as amorphization. This study provides critical insights into mechanics of $CH_3NH_3PbI_3$ perovskites, and offers that TB engineering is a promising strategy to design mechanically robust hybrid organic-inorganic perovskites-based device systems.

**Keywords:** $CH_3NH_3PbI_3$ perovskite; Twin boundary; Mechanical detwinning; Ductile characteristics; Molecular dynamic simulations

*Corresponding Emails: zhangzs@xmu.edu.cn, jianyang@xmu.edu.cn

# 1. Introduction

Hybrid organic-inorganic perovskites (HOIPs) have been identified to show structural diversity, unique optical and electronic properties, such as long carrier diffusion lengths, high carrier mobility, tunable and small bandgaps, strong extinction coefficients, strong photoluminescence performance, and so on[1-5]. As a consequence, they have a large variety of important practical applications, for example, their utilization as optoelectronic and photonic devices, such as solar cells, photodetectors, light-emitting diodes, field-effect transistors, waveguides, as well as nanolasers[2, 6-8].

As a prototype of HOIPs, methylammonium lead triiodide ($CH_3NH_3PbI_3$ or $MAPbI_3$) is one of the most promising energy-related materials for utilization in the next-generation low-cost, low-temperature, and solution-processable photovoltaic and optoelectronic devices with high conversion efficient recover[9, 10]. Compared to its inorganic counterparts, as a result of the presence of organic ions, $MAPbI_3$ is mechanically soft nature of the hybrid framework and shows better performance in response to mechanical loading[11], providing broad prospect for further potential applications in flexible and soft-wearable functional devices[12, 13].

Because of those important practical applications, there have been a number of pioneering both experimental and theoretical studies examining the mechanical characteristics of $MAPbI_3$ perovskite[11, 14-19] that determine the service lifetime and reliability of electronic devices. Using inelastic neutron scattering and brillouin scattering, it was measured that the bulk modulus of single crystal of $MAPbI_3$ perovskite is $14.8 \pm 1.7$ GPa at 340 K[20], and the mechanical stability can be improved by doping proper concentration Cs[21]. Using nanoindentation technique, it was reported that elastic modulus and mechanical hardness of single crystal of $MAPbI_3$ perovskite varies from around 10.4 - 20.0 GPa and about 0.47 - 1.0 GPa[18, 19, 22-24], respectively, depending on the measured crystalline faces, sample quality, and so on. Using density functional theory (DFT) and molecular dynamics (MD) simulations, Young's modulus of single crystal of MAPbI$_3$ perovskite was determined to vary from 12.3 - 22.2 GPa, and MAPbI$_3$ perovskite is more mechanically robust than Cl- and Br-based perovskite[11, 14-17]. Moreover, its bulk modulus was calculated to vary from 15.6 -18.5 GPa[15, 16].

As is known, however, due to its fabrication mainly by process-solution method, MAPbI$_3$ perovskite is invariably polycrystalline with intrinsic structural defects such as vacancy, grain boundary (GB)[25, 26]. It was theoretically and experimentally studied that structural defects such as GBs can show either positive or negative effects on the physical properties and performance of MAPbI$_3$-based device[27-29]. Such conflicting results are primarily attributed to the fact that the tested polycrystalline samples show very differences in the microscopic structure and morphology, including grain size, density of GBs, twin boundary (TB) and so forth[11, 30-32].

In mechanics, polycrystalline MAPbI$_3$ perovskites are mechanically weaker than its single crystal counterpart. Using scanning probe microscopy technique, the elastic modulus of polycrystalline MAPbI$_3$ perovskite films was determined to be around 9.00 GPa[33]. As a result of reduction in GB defects, polycrystalline perovskite thin films containing large grain sizes show high overall fracture resistance and mechanical stability [34-36]. Moreover, the fracture energy ($G_c$) of polycrystals ranges from around 0.41 ± 0.17 - 1.14 ± 0.24 J/m$^2$ that is much lower than that of single crystal (2.7 J.m$^{-2}$) [22, 34], and the microscope residual stress can induce GBs' microcrack propagation in the polycrystalline thin films[22]. Using MD simulations, it was shown that polycrystalline MAPbI$_3$ perovskite exhibits Young's modulus of around 5.08 - 7.89 GPa, but higher ductility than single-crystal [11]. Such apparent reduction in the elastic properties can be mainly attributed to the complex microstructures at the polycrystalline grain boundaries (GBs) and finite crystalline grain size[37].

To the best of the our knowledge, however, the critical role of GBs on the mechanical characteristics of MAPbI$_3$ perovskite remains largely unexplored yet, particularly for the role of TBs, known as a special type of two-dimensional (2D) structural defects, on its mechanics. Previous reports show the existence of TBs, for example, Σ3(111) TB, positively enhances the physical properties of MAPbI$_3$ perovskite and the performance of solar cells[30, 38]. Understanding the influence of TBs on the mechanical stability and mechanical properties of MAPbI$_3$ perovskite is of crucial for improving performance of its solar cells with good mechanical flexibility and robustness. To this end, in this study, a variety of distinct TB structures in MAPbI$_3$ perovskite are constructed on the basis of coincidence site lattice (CSL) method[39, 40] and their tensile mechanical characteristics are for the first time examined using classic MD simulations.

## 2. Models and Methods

### 2.1 Molecular Models

MAPbI$_3$ perovskite structure is a basic building block of organic-inorganic perovskite family. Structurally, MAPbI$_3$ perovskite is a class of materials with similar crystal structure as calcium titanate (CaTiO$_3$), showing an ABX$_3$ configuration, where the A, B and X sites are occupied by small organic cation, divalent group element, and halogen, respectively[41]. Here, the A is CH$_3$NH$_3^+$ (MA$^+$), the B is cation of Pb$^{2+}$, and the X is halogen I$^-$. MAPbI$_3$ perovskite structure is cubic lattice with space group of Pm3m, and the lattice constants of *a*, *b* and *c* are equal to 8.8 Å[42, 43].

In this work, ten systems including pristine and nine bicrystalline structures containing different TB defects are taken into considerations, as shown in Table 1 and Figure 1. There are differences in microstructures between single crystal and bicrystals. Bicrystalline MAPbI$_3$ perovskite are generated from CSL model. Using CSL model, two grains are twisted to form a definite angle until their boundary structures coincided to generate bicrystalline models. Geometries of CSL TBs are characterized by $\Sigma$ value that is defined as the inverse of coincident site density. Large $\Sigma$ values indicate random boundary and low-periodicity, while low $\Sigma$ values represent characteristic high-periodicity and low interface energy [44]. Reciprocal planar coincidence density of crystal lattice ($\Sigma$) is expressed as

$$\Sigma = n^2 + m^2(u^2 + v^2 + w^2) \tag{1}$$

where *m* and *n* are non-reciprocal integers. Moreover, [*m*, *n*] represents the coordinates of coincident lattice. The [*u*, *v*, *w*] are the common axis that TB rotates around, the [001] is the rotated axis in this work.

Alternately, to quantitatively characterize the bicrystals of MAPbI$_3$ structures, tilt TB angle is defined. Here, the TB angle varies from around 11.3° - 38.7°. The tilt TB angles in bicrystalline structures are calculated as follows:

$$\theta = 2\tan^{-1}\left(\frac{m}{n}\sqrt{u^2 + v^2 + w^2}\right) \tag{2}$$

Apparently, as shown Figure 1, bicrystals with different tilt TB angles show very different microstructures of TBs. Moreover, TB energy ($\sigma_{TB}$) of bicrystalline MAPbI$_3$ structures with respect to bulk phase is defined as:

$$\sigma_{TB} = \frac{E_{BI} - E_{Bulk}}{2A_{TB}} \tag{3}$$

where $E_{BI}$ and $E_{Bulk}$ are total potential energies of bicrystalline and bulk MAPbI$_3$ structures, and $A_{TB}$ is the lateral area perpendicular to loading direction in cells with TB structures.

**2.2 Tension MD Simulations**

Initially, all MAPbI$_3$ samples are quasi-statically optimized to a configuration with local minimum energy via conjugate gradient method, where the stopping tolerances of energy and force are $1.0 \times 10^{-4}$ Kcal/mol and $1.0 \times 10^{-4}$ Kcal/(mol·Å), respectively. Afterwards, MD relaxations are carried out with

$6.0 \times 10^6$ timesteps for pristine crystal and $1.5 \times 10^7$ timesteps for bicrystals at given temperature and standard atmosphere under NPT (constant number of particles, constant pressure, and constant temperature) ensemble. Posterior to full structural relaxations, the samples are uniaxially stretched with a constant engineering straining rate of $10^8$/s along one orthogonal direction that is perpendicular to TB plane. Both Nosé-Hoover thermostat and barostat techniques with damping times of 100 and 1000 timesteps are employed to control the system temperature and non-loading directional pressures, respectively. The velocity-Verlet algorithm with a timestep of 0.1 fs is implemented to integrate the Newton's equations in the MD runs. Periodic boundary conditions (PBCs) are applied in all three orthogonal directions. Initial velocities of all atoms in the MAPbI$_3$ samples are assigned on the basis of Gaussian distribution at given temperatures. By using the virial definition of stress, the atomic stress is achieved through collection of the forces on atoms during MD runs. Atomic stress in all MAPbI$_3$ samples are averaged over 1000 timesteps to eliminate the thermal fluctuations. To statistically capture the mechanical properties, each sample with five different initial Gaussian distributions of atomic velocities is considered. All the MD runs are implemented using the Large-scale Atomic-Molecular Massively Parallel Simulator (LAMMPS) package[45].

**2.3 Forcefield for Modelling MAPbI$_3$**

To mimic the atomic interactions in the MAPbI$_3$ perovskite systems, including inorganic-inorganic, organic-organic and organic-inorganic interactions, the classical model potential for hybrid perovskites (MYP)[46] is employed. The MYP forcefield consists of Buckingham, Coulombic and Lennard-Jones (LJ) potential terms. The atomic interactions between the inorganic parts (Pb and I) are described by a combination of Buckingham and Coulombic potentials with mathematical expression as follows[47]:

$$U_{ij} = A \exp(-r_{ij}/\rho) - \frac{C_6}{r_{ij}^6} + \frac{q_i q_j}{r_{ij}} \tag{4}$$

Within this expression, the exponential item indicates the short-ranged repulsive overlap interactions between pairs of ions, while the second item represents long-ranged attractive van der Waals (vdW) interactions between ions. Those two terms represent the Buckingham forcefield. The last item indicates the Coulombic interaction between pairs of charged ions ($q$)[48]. Besides, the sum of 12-6 Lennard-Jones (12-6 LJ) and Coulombic potentials are utilized to describe interactions of MA$^+$ and inorganic parts[49], with mathematical expression as follows:

$$U(r_{ij}) = 4\epsilon_n \left[ \left(\frac{\sigma_{ij}}{r_{ij}}\right)^{12} - \left(\frac{\sigma_{ij}}{r_{ij}}\right)^6 \right] + \frac{q_i q_j}{4\pi\epsilon_0 r_{ij}} \tag{5}$$

where $r_{ij}$ is the distance between interacting atoms $i$ and $j$. $\epsilon_n$ and $\sigma$ are the 12-LJ parameters, representing the LJ potential well depth with the finite distance $\sigma$ at which the interatomic potential equals to zero. $q$ is ionic charge. Moreover, the standard GAFF forcefield[50, 51] is employed to describe the interactions between organic MA$^+$ ion.

## 3 Results and Discussion

### 3.1 Distortion and Energetics in TB Structures

The chemical and physical properties of a polycrystalline material is dominted by their microstrcutural configurations and chemistry of GBs, morphology of GB networks and crystalline grain size. The properties of GBs that detemine the properties of polycrstals can be specifically characterzied by GB misorientation angle and GB plane. As is seen from Figure 1, as a result of finite mirror twisting of two MAPbI$_3$ perovskite crystals to form bicrystals, unusual non-4-membered rings composed of bielemental inorganic polar Pb-I bonds appear in the TB structures, which are characteristic defects in GBs of MAPbI$_3$ perovskite crystals. Similar to bielemental 2D MoS$_2$ sheets[52, 53], MAPbI$_3$ perovskite crystals show diversity in the defective rings of TB structures. Bicrystal with different tilt angles of TB exhibits significant difference in the defective structures in the TB region. For example, TB structures with small tilt angles of 11.3°, 14.0° and 18.4° are mainly composed of pairs with Pb-polar 5|5-membered rings, whereas that with tilt angle of 21.8° is structurally characterized by I-polar 5 + 5|5 membered rings. Note that those 5|5-membered rings are in different geometry. In contrast, defective rings in TB structures with large tilt angle varying from 30.9 - 38.7° are composed of large number of Pb-I bonds, resulting in their large area of rings. Due to the appearance of those defective non-4-membered rings, MAPbI$_3$ perovskite crystals are structurally distorted from the square geometry to parallelogram motif in the inorganic framework, particularly in the vicinity of TB structures. Such distortion causes deviation in critical bond Pb-I-Pb angles, which is about 170-176.5°[54, 55] generating intrinsic excess free energy in MAPbI$_3$ perovskite bicrystals.

Commonly, formation energies of GBs quantitatively indicates the structural stability of polycrystals. In this work, potential energies of a set of TBs for MAPbI$_3$ perovskite structure are first evaluated at finite temperatures. Figure 2 shows the variations in the energies of 9 distinct TBs with temperature varying from 123.15-350 K. As is expected, TB energy is highly temperature dependent. As a consequence of structural distortions induced by thermal fluctuations, with the increase of the temperature, all bicrystals of MAPbI$_3$ perovskite show monotonic rising in TB energy and the rising becomes less pronounced. Within the external temperature conditions, MAPbI$_3$ perovskite bicrystals show TB energy of around 1.0 - 3.0 J/m$^2$, depending on the tilt angle of TB, which is higher than that of guest-free pure inorganic perovskites[56]. As is known, structures with lower GB energy are more structurally stable[57]. Therefore, this explains the fact that polycrystalline MAPbI$_3$ perovskites are less structurally stable than polycrystalline inorganic perovskites. By comparison, it is observed that, with temperature varying from 123.15 - 350 K, TB structure of tilt angle of 26.6° shows the highest TB energy, with a range of around 2.1 - 2.9 J/m$^2$, and TBs with tilt angles of 11.3°, 14.0°, 18.4°, 30.9° and 36.9° yield intermediate TB energies of about 1.1 - 2.4 J/m$^2$. Whereas, TBs with tilt angles of 21.8° and 38.7° present the lowest TB energies, varying from 1.0 - 1.7 J/m$^2$, representing the energetically most stable TB configurations at finite temperatures.

### 3.2 Uniaxial Tensile Detwinning Responses

As is known, polycrystalline MAPbI$_3$ thin films based optoelectronic and photonic devices, for example, solar cells, actually work at elevated temperatures of around 343 - 353 K[54, 58], and polycrystalline MAPbI$_3$ thin films are inevitably deformed due to various external loads. Mechanical stability and performance of polycystalline MAPbI$_3$ perovskite-based devices are greatly dictated by the mechanical properties of their microscopic GBs and grain size. Therefore, the tensile characteristics of above-constructed TBs are examined at finite temperatures of $\leqslant$ 350K , as well as their single crystal counterpart. Figure 3 presents the uniaxial tensile stress-strain curves of single crystal and bicrystals with tilt TB angle varying from 11.3 - 38.7° at finite temperature of 123.15 - 350 K. Obviously, defect-free single crystal is more mechanically robust structure than bicrystals that contain defects in the TB region. According to the feature of stretching curves, three or four loading stages can be roughly identified, depending on the type of TB structure, which is insensitive to the external temperature; TB structure with tilt angle of 26.6° show four distinct loading stages, while the others present three distinct loading stages, as well as the single crystal counterpart.

For all the structures mentioned, the first loading stage is described by the initial linear tensile-strain curve within very limit strain regime, representing linear elastic deformation. The second loading stage of MAPbI$_3$ structures, except for the TB structure with tilt angle of 26.6°, is mainly characterized by that stretching stresses are nonlinearly increased up to their maximum values, namely, rising in stretching stresses becomes less pronounced with increasing tension strain, demonstrating strain-softening behavior. For the case of TB structure with tilt angle of 26.6°, the second loading stage is described by nonlinear increase in the tensile stress up to the second highest peaks at strain of around 0.025 - 0.05, whereas its third loading stage corresponds to nonlinear rising in loading stress from the second highest peak to the highest peak. The significant drop of tensile stress posterior to the second loading stage indicates occurrence of dramatic structural transformation. According to the concavity of variation in the potential energy with strain, the nonlinear loading responses of all structures are distinguished as plastic deformation. The last loading stage of $MAPbI_3$ single crystal is mainly described by gradual reduction in the tensile stress. Whereas for $MAPbI_3$ bicrystals, it is primarily characterized by that loading stresses are largely fluctuated with increasing tension strain. The final loading stage of all structures correspond to post-failure deformation.

Based on the characteristics of tensile loading responses, those investigated $MAPbI_3$ structures can be classified into four types. Type I is represented by single-crystal and TB structures with tilt angle of 11.3°, 14.0° and 18.4° and 21.8° that show reduction of loading stress with small fluctuation in the final stage. Remarkably, reduction rate of the loading stress for single crystal is more significant than that for the TB structures, indicative of higher mechanical ductility of polycrystals than that of single crystal, in agreement with previous reports[11]. Type II consists of TB structures of 30.9°, 33.7° and 38.7° tilt angles, and their last loading stages are characterized by that a sudden deep drop of tensile stresses initially occur, suggesting occurrence of large structural change, after which high-amplitude fluctuation of tensile stress follows. TB structure of 36.9° tilt angle belongs to the structural Type III, and the final loading stage is described by sudden deep drop of stretching stress, followed by rising trend of loading stress that is in contrast to the case of Type I and II. Type IV only contains TB configuration with 26.6° tilt angle, and it is primarily characterized by double sharp loading peaks in the intermediate loading stages II and III.

**3.3 Tensile Stiffness**

Tensile stiffness such as Young's modulus ($E$) are extracted directly from the tensile stress-strain curves of Figure 3. Note that Young's modulus is defined by the ratio of tensile stress ($\sigma$) to strain ($\varepsilon$) (within the initial linear elastic regime of 0.00-0.005). Figure 4 plots the variation in the tensile Young's modulus of single crystal and bicrystals of TB tilt angle varying from 11.3 - 38.7° with temperature. For pristine crystal, it is observed from Figure 4a that its Young's modulus, depending on the temperature, ranges from around 15.0 - 19.2 GPa. Previous experimental measurements and numerical simulations showed that Young's modulus of single crystal of MAPbI$_3$ perovskite varies from 10.4 - 22.2 GPa[11, 15-19, 23, 59, 60], relying on the measurement and numerical techniques, as well as temperature and loading crystallographic orientations. Apparently, it is inferred that our MD calculations are in good agreement with those above reported results, verifying the adopted MYP forcefield for mechanical prediction of MAPbI$_3$ perovskite. Interestingly, with increasing temperature from 123.15 - 350 K, it is identified non-monotonic change in the Young's modulus of pristine MAPbI$_3$ perovskite, namely, there is no apparent reduction in the Young's modulus with rising temperature, which is against thermal softening mechanism. This is because, at different temperature conditions, MAPbI$_3$ perovskite forms distinct mechanically robust crystalline phase structures, including orthorhombic, tetragonal and cubic phases[61], for example, the critical temperature (162 K) of orthorhombic-tetragonal crystalline, cubic phase at temperature of around 327 K[42, 58, 62]. Using nanoindentation technique, it is reported that tetragonal phase of MAPbI$_3$ perovskite show Young's modulus of around 10.4 - 10.7 GPa[19], depending on the measured crystallographic orientation, 14.3 GPa on (110) face and 14.0 GPa on (112) face at room temperature with tetragonal (pseudo cubic) phase[18].

For bicrystals, as shown in Figure 4b-j, their tensile Young's moduli are significantly lower than that of pristine counterpart. This results from the fact that there are defects in the TB region of bicrystals. Apparently, the Young's modulus of bicrystals is greatly type of TB structure and temperature dependent. TB structures of 11.3°, 14.0°, 18.4° and 21.8° tilt angles belonging to Type I exhibit Young's modulus of around 7.5 - 12.8 GPa. Intriguingly, at critical temperature of around 180 - 200 K, there is an apparent deep drop in the Young's modulus for TB structures with 14.0°, 18.4° and 21.8° tilt angles, whereas above which, their Young's moduli are insensitive to the temperature. In contrast, for TB structure with 11.3° tilt angle, its Young's modulus is decreased with the increasing temperature. With regard to TB structures of Type II containing 30.9°, 33.7° and 38.7°, it is observed that their Young's moduli are initially monotonically reduced as the temperature increased from 123.15 - 200 K, but then remain unchanged with further increasing temperature. For the case of TB structure with tilt angle of 36.9° that belongs to Type III, it is identified a monotonic reduction in tensile Young's modulus with the increasing temperature, following the thermal softening mechanism. However, tensile Young's modulus of TB structure of 26.6° tilt angle is found to be insensitive to the temperature, similar to the case of pristine crystal. In terms of the value of tensile Young's modulus at low/high temperature condition, TB structure of 33.7/38.7° tilt angle is the most mechanically robust configuration, while TB structure with 30.9° tilt angle is the least mechanically robust one.

**3.4 Tensile Detwinning Strength**

Tensile detwinning strength can be characterized by the maximum tensile strength extracted from the loading curves in Figure 3. Figure 5 shows variations in the maximum tensile strength of pristine crystal and nine distinct bicrystals with temperature. As is seen from Figure 5a, pristine crystal show the maximum tensile strength of > 0.45 GPa at temperature varying from 123.15 - 350 K. Interestingly, there is rising tendency in the maximum tensile strength with increasing temperature, breaking the thermally activated weakening mechanism. Such thermally-induced strengthening can be understood by the occurrence of phase transition of MAPbI$_3$ perovskite [61].

With regard to bicrystals, as a consequence of weak defects located at the TB region, they exhibit lower maximum tensile strength than that of single crystal (Figure 5a). All samples present monotonic reduction in the maximum tensile strength with increasing temperature as a result of thermal weakening at the TB region, in sharp contrast to the case of single crystal counterpart. The reduction in maximum tensile strength of Types I, II and IV TB structures becomes less pronounced with the increasing of temperature, whereas for the Type III TB structure of tilt angle of 36.9°, it becomes more significant with rising temperature. Moreover, TB structure with different tilt angle presents different sensitivity to the temperature. By contrast, TB structure with 36.9/14.0° tilt angle is the most/least sensitive configuration in the maximum tensile strength to temperature, with reduction of around 0.3/0.13 GPa as temperature is increased from 123.15 - 350 K. For a given temperature, in terms of the value of maximum tensile strength, TB structure with 36.9° tilt angle is identified to be least mechanical robust configuration in all the structures. Note that both tensile Young's modulus and the maximum tensile strength are not positively correlated with the TB energy. This is because mechanical properties of MAPbI3 perovskite composed of and entrapped organic MA+ are determined not only by bonding properties of Pb-I inorganic octahedral cage frameworks, but also by the non-bonded interactions between octahedral cages and guest ions of MA+[11], differing from the case of other inorganic materials[63, 64].

## 3.5 Tensile Ductile Detwinning Mechanisms

To in-depth uncover the deformation mechanisms underlying type of TBs dependent uniaxial loading responses, developments of molecular structures of all bicrystalline MAPbI$_3$ perovskites subjected to tension are recorded. As discussed above, bicrystals with different TB tilt angles can be divided into four types. Figure 6 presents the four representative mechanical loading curves and their corresponding snapshots of TB structures subjected to critical tension strains as indicated by the solid circles in the curves. For clarification, organic ions of MA$^+$ are removed and inorganic (Pb and I) atoms are colored on the basis of values of atomic shear strain. At relaxed state (zero tension strain), all investigated TB structures are configurational stable. As indicated by both loading curves and snapshots in Figure 6, there are distinct stretching ductile deformation mechanisms between the four types of TB structures.

For bicrystals consisting of Type I or II TB structure, as indicated by Figure 6e and f, the loading stages I and II are mainly dominated by relative uniform distribution of structural shear deformation. There are no significant crystalline structural changes, explaining the relative smooth changes of tensile stresses. In the loading stage III, however, tension-induced shear strain is locally concentrated at TB regions. As a result, TBs are destructed with increasing strain. By comparison, the strain-induced destruction of TBs of Type I is described by slow change in TB configurations, explaining the flow characteristic of loading stress. Whereas for TBs of Type II, it is mainly characterized by a rapid disordering in TB configurations, elucidating the sudden deep drop of loading stress in the tensile curve. This can be considered as characteristic brittle failure. Interestingly, those destructions of TBs do not lead to rapid separation of the two contacting crystals, as well as strain release in the bulk region, indicating that the disordering TBs are still able to carry load as a result of reformation of Pd-I bonds and non-bonded interactions. This is the molecular origin of the super tensile ductile deformability.

With regard to bicrystal containing Type III TB structure, as shown in the first two snapshots of Figure 6g, the first two loading stages are characterized by that stretching load-induced shear strain is primarily carried on the TB region and in the vicinity of TB structure. Consequently, as the load is critically imposed, ordered TB configurations and its neighboring defect-free structures are rapidly destructed (snapshot #3 of Figure 6g), reflecting brittle failure, analogous to that in Type II TBs. Whereafter, as a result of the large-area failure, there is rising tendency in the loading stress with increasing tension strain. And the super ductile deformation is contributed by elongation of large-area as-destructed structures.

As for bicrystal composed of Type IV TB structure subjected to uniaxial tension, shear strain is non-uniformly distributed, with tension-induced shear strain on localized region and some neighboring defect-free Pb-I cages of TB structure. As the tension strain reaches to critical value, bond dissociation of Pb-I occurs at the locally load carried regions, resulting in the first apparent reduction and oscillation of loading stress (snapshot #2 of Figure 6h). With augment of uniaxial tension strain, stress concentration becomes well distributed on the whole TB structure, leading to re-rising in significant global loading stress. Once the bicrystal is critically uniaxially elongated, however, highly strained TBs are catastrophically destructed by rapid dissociation and reformation of Pb-I bonds, producing amorphization at the TB region. Such rapid destruction of TB structure causes sudden deep drop of loading stress (Figure 6d). This reveals that Type IV TB structure shows characteristic brittle failure in this loading phase. Afterwards, its deformation mechanism is similar with that of Types I and II, with elongation of amorphized TB structures.

In a nutshell, compared with single crystal MAPbI$_3$ perovskite, bicrystals containing TB structures are more easy deformed and failed, similar to the reinforced polymers and glassy ceramics in fracture mechanism[11]. However, they exhibit unique tensile ductile mechanisms, depending on the type of contained TB structure, and are different from that of single crystal. As is indicated, molecular structures of TBs play critical roles in structural stability and mechanical performance of polycrystalline MAPbI$_3$ perovskites that are critical to performance and lifetime of MAPbI$_3$ perovskite-based device systems.

## 4. Conclusions

In summary, in this work, tensile mechanical detwinning properties of MAPbI$_3$ perovskites containing nine distinct TB structures are for the first time investigated using classic MD simulations. It is revealed that tensile properties such as Young's modulus, maximum detwinning strength, as well as the formation energy of TBs are greatly temperature and tilt angle of TBs dependent. There is anti-correlation between Young's modulus/maximum detwinning strength and temperature, but complex correlation between structural/tensile properties (formation energy of TBs/Young's modulus and maximum tensile strength) and tilt angle of TB structure.

All investigated TB-contained MAPbI$_3$ show superior tensile detwinning ductility over single crystal counterpart. There are four types of TB structures classified from the tensile detwinning characteristics. Similar to the bulk counterpart, Type I structure containing TBs with tilt angles of 11.3°, 14.0°, 18.4° and 21.8° exhibit smooth ductile flow stress-strain curves because of the relatively uniform load carry and gradual amorphization of TB structure. Type II structure including TB tilt angle of 30.9°, 33.7° and 38.7° exhibit sudden drop of loading stress followed by ductile flow behavior, originating from rapid amorphization in limit TB structure, however, Type III structure consisting of TB tilt angle of 36.9° is characterized by rising trend in loading stress prior to sudden stress drop, resulting from large-area amorphization in TB and its neighboring regions. Intriguingly, Type IV structure represented by TB tilt angle of 26.6° shows double peaks in the tensile curve followed by ductile flow stress characteristic, resulting from stress carry of localization-to-globalization at TB region and amorphization of TB structure. This work provides new insights into mechanical characteristics of MAPbI$_3$ containing TBs, and HOIPs can be fabricated with enhanced structural and mechanical stability via TB engineering for utilization in various photovoltaic industries and soft-wearable devices.

## Acknowledgments

This work is financially supported by the National Natural Science Foundation of China (Grant Nos. 11772278 and 11904300), the Jiangxi Provincial Outstanding Young Talents Program (Grant No. 20192BCBL23029). Y. Yu and Z. Xu from Information and Network Center of Xiamen University for the help with the high-performance computer.

**Tables, Figs and captions**

Table 1. Molecular information of bicrystalline MAPbI$_3$ models

| Model | Tilt angle (°) | No. A, B and X groups | Dimensions (nm$^3$) |
|---|---|---|---|
| Single-crystal | 0° | 1000, 125, 375 | 3.14 × 3.11 × 3.19 |
| Σ103(510) | 11.3° | 4800, 624, 1848 | 12.9 × 3.21 × 3.78 |
| Σ67(410) | 14.0° | 9600, 1248, 3696 | 15.6 × 5.20 × 3.78 |
| Σ39(310) | 18.4° | 7296, 960, 2832 | 15.9 × 3.98 × 3.78 |
| Σ7(520) | 21.8° | 5376, 720, 2064 | 13.6 × 3.39 × 3.78 |
| Σ5(210) | 26.6° | 8064, 1080, 3240 | 16.9 × 4.23 × 3.78 |
| Σ7(530) | 30.9° | 6144, 816, 2424 | 14.7 × 3.67 × 3.78 |
| Σ3(320) | 33.7° | 9600, 1248, 3696 | 18.2 × 4.54 × 3.78 |
| Σ5(430) | 36.9° | 7104, 912, 2664 | 18.9 × 3.15 × 3.78 |
| Σ9(540) | 38.7° | 7680, 984, 2856 | 16.1 × 4.03 × 3.78 |

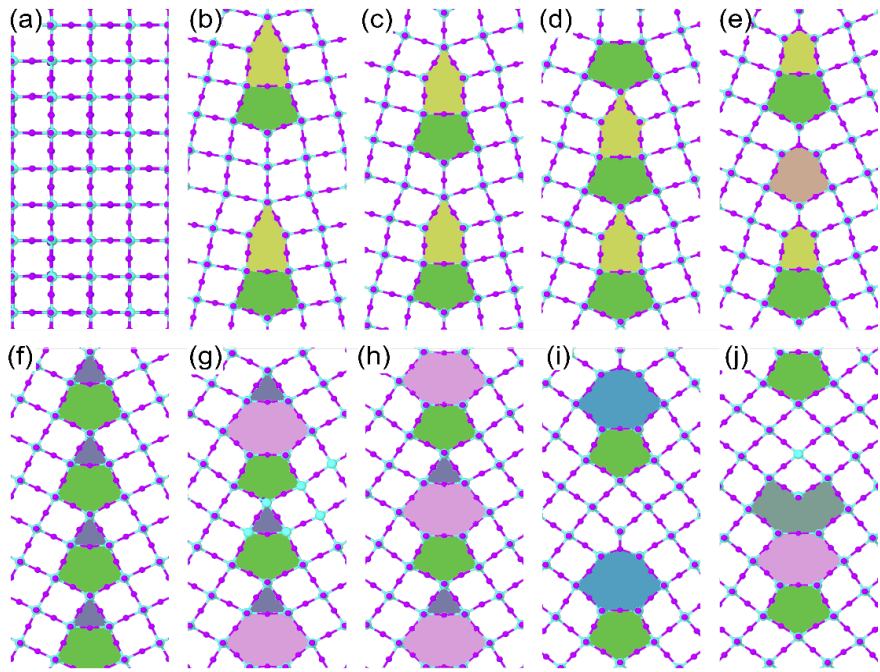

Figure 1 Molecular structure of MAPbI$_3$ perovskite frameworks. (a) Single crystal structure composed of Pb-I inorganic octahedral cage framework. (b) - (j) Bicrystals containing twin boundaries (TBs) with tilt angle varying from 11.3 - 38.7°, respectively. To highlight the structures of TBs, irregular polygons in TB region are differently colored.

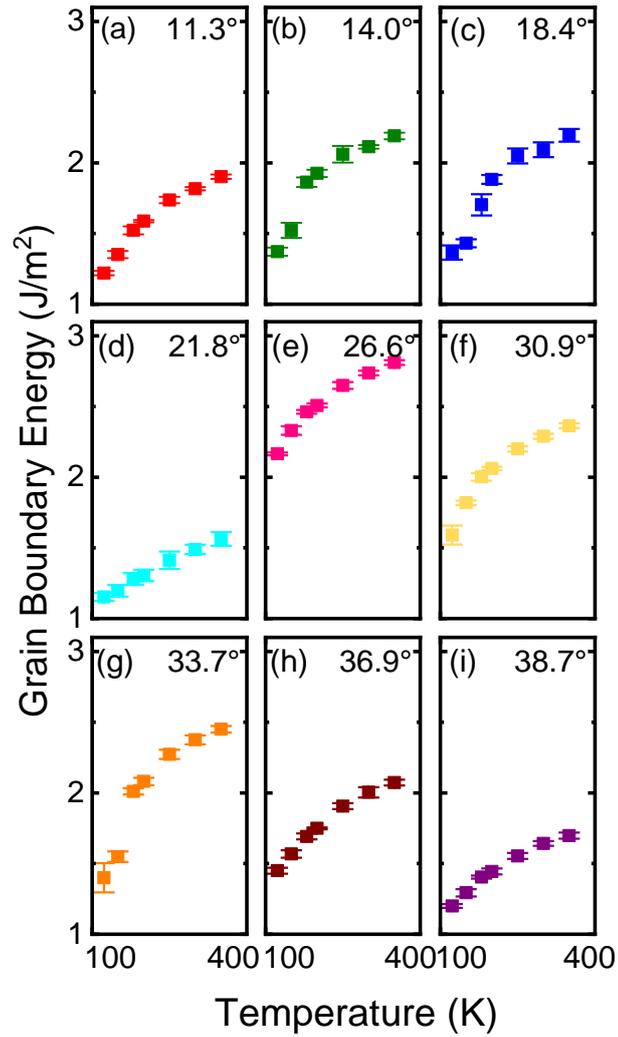

Figure 2 Energetics of twin boundaries (TBs) in bicrystalline MAPbI$_3$ perovskites. (a) - (i) Variations in the formation energy of TBs of tilt angle varying from 11.3 - 38.7° with temperature from 123.15 - 350 K, respectively.

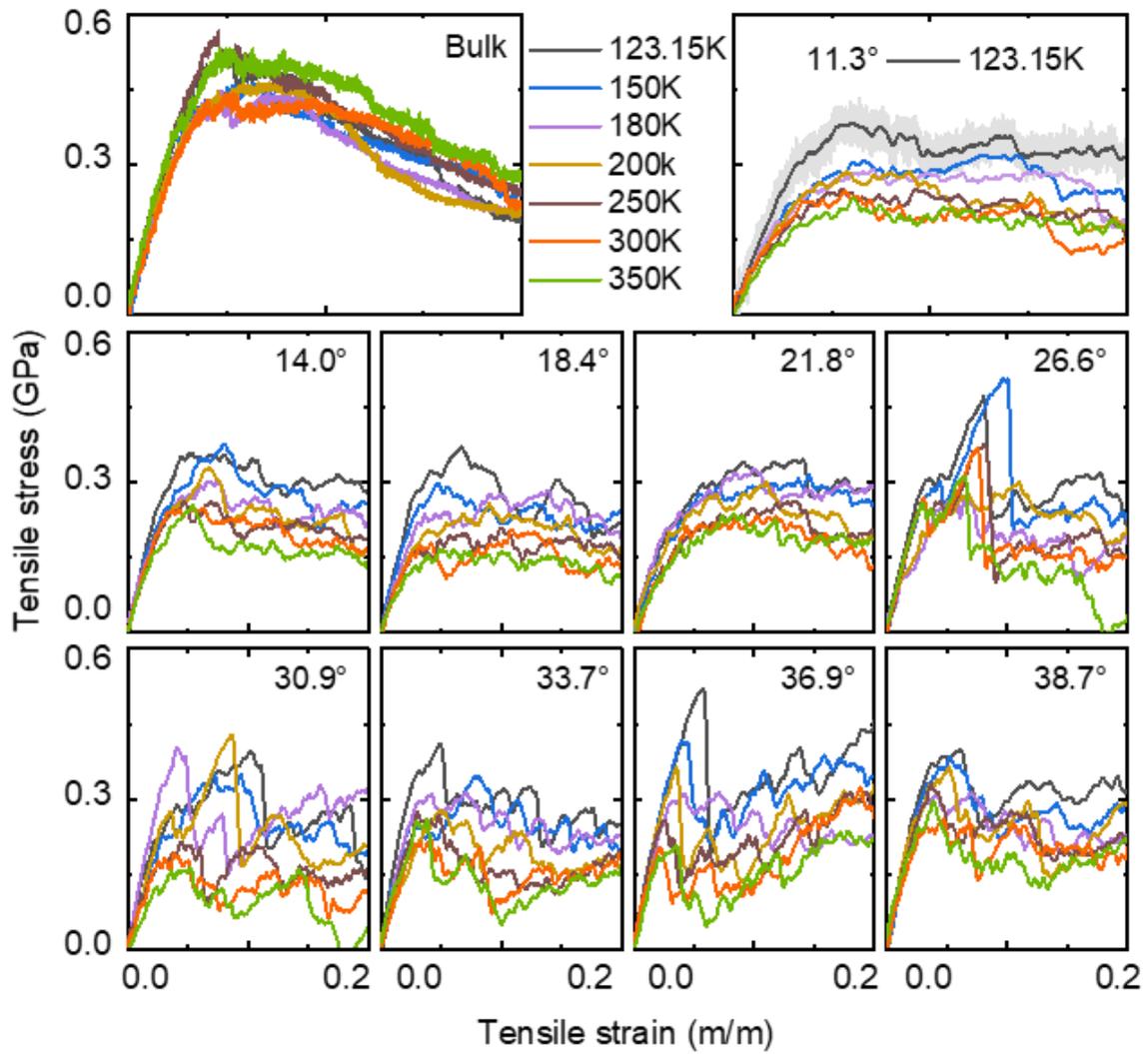

Figure 3 Uniaxial tensile loading stress-strain curves of single crystal and bicrystals containing twin boundaries (TBs) with tilt angle of 11.3 - 38.7° of MAPbI$_3$ perovskites at temperature varying from 123.15 - 350 K.

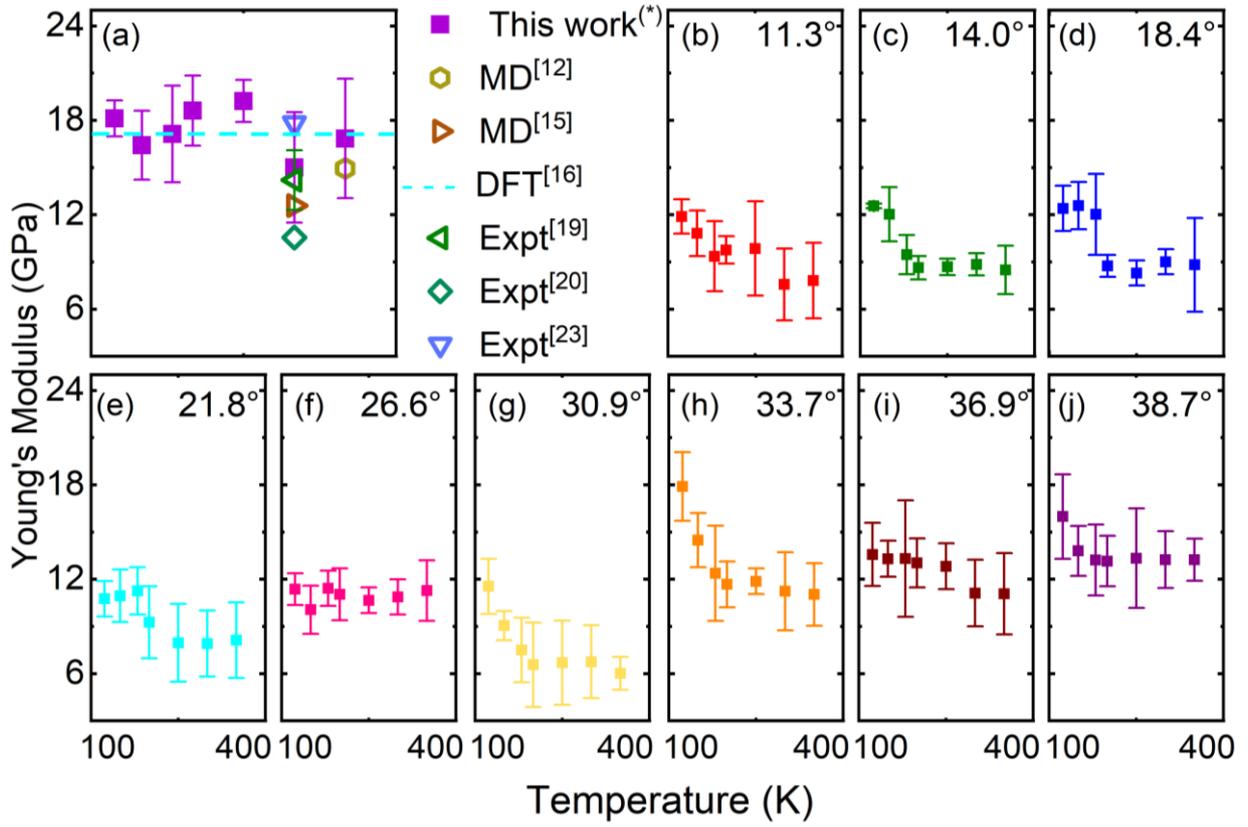

Figure 4. Tensile stiffness of MAPbI$_3$ perovskites. (a) Young's modulus of single crystal of MAPbI$_3$ perovskite from our MD predictions at temperatures of 123.15 - 350 K, as well as the experimental measurements[18] [19] [22], MD and DFT calculations[11] [14, 15]. (b) - (j) Young's modulus of bicrystals of MAPbI$_3$ perovskite composed of twin boundaries (TB) of tilt angle varying from 11.3 - 38.7° as a function of temperature from 123.15 - 350 K, respectively.

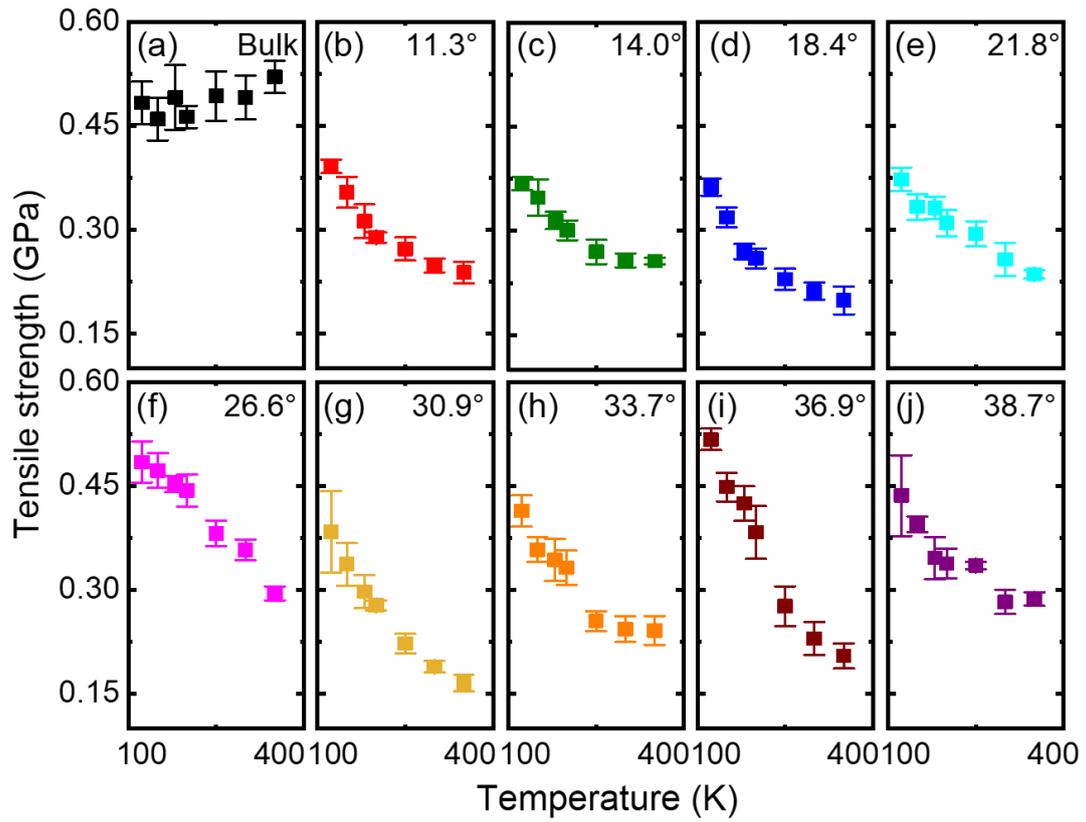

Figure 5 Tensile strength of MAPbI$_3$ perovskites. (a) Maximum tensile strength of single crystal of MAPbI$_3$ perovskite as a function of temperature from 123.15 - 350 K. (b) - (j) Variations in the maximum tensile detwinning strength of bicrystals consisting of twin boundaries (TBs) of tilt angle changing from 11.3 - 38.7° with temperature varying from 123.15 - 350 K.

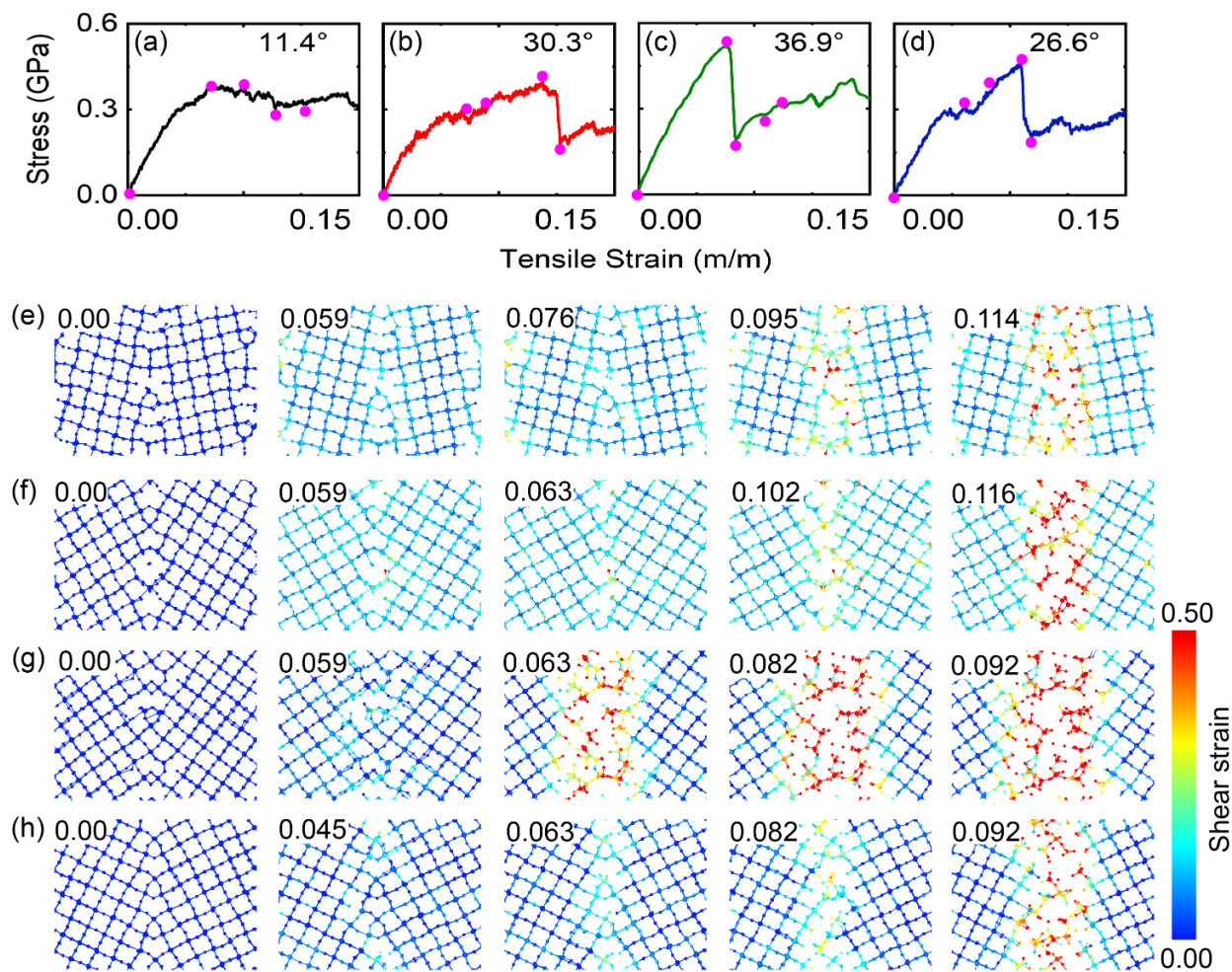

Figure 6 Uniaxial tensile detwinning ductile deformation mechanisms of $MAPbI_3$ perovskites. (a) - (d) Tensile stress-strain curves of bicrystals of $MAPbI_3$ perovskites containing twin boundaries (TBs) with tilt angles of 11.3°, 30.3° and 36.9° and 26.6°, respectively. Those four TBs are representative Types I, II, III and IV TB structures. (e) - (f) Microstructural snapshots of TB structures with tilt angles of 11.3°, 30.3°, 36.9° and 26.6° at critical strains that are indicated by the solid purple circles in the loading curves, respectively. For clarification, organic $MA^+$ ions are removed and the color code is based on the values of atomic shear strains.

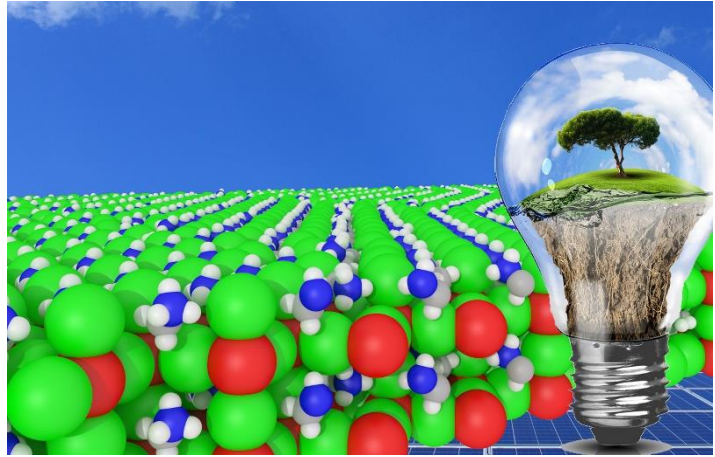

TOC Figure